\documentclass[10pt,conference]{IEEEtran}
\IEEEoverridecommandlockouts
\usepackage{cite}
\usepackage{amsmath,amssymb,amsfonts}
\usepackage{algorithmic}
\usepackage{graphicx}
\usepackage{textcomp}

\usepackage{cite}
\usepackage{amsmath,amssymb,amsfonts}
\usepackage{algorithmic}
\usepackage{graphicx}
\usepackage{textcomp}

\usepackage{listings}
\usepackage{array} 
\usepackage{ragged2e}
\usepackage{enumitem}
\usepackage{url}
\usepackage{wrapfig}
\usepackage{subcaption}
\usepackage{caption}
\usepackage{multirow}
\usepackage{flushend}
\usepackage{tikz}
\usepackage{censor}
\usepackage[most]{tcolorbox}
\usepackage{soul}

\usepackage{textcomp}
\usepackage{amsmath}
\usepackage{array, multirow, bigdelim, makecell, booktabs} 
\usepackage{xargs}   
\tcbuselibrary{theorems}
\newtcolorbox{mybox}[1]{colback=white!5!white,colframe=gray!45!white, title =#1,coltitle=black!20!black}

\colorlet{punct}{red!60!black}
\definecolor{background}{HTML}{EEEEEE}
\definecolor{delim}{RGB}{20,105,176}
\colorlet{numb}{magenta!60!black}

\definecolor{plantucolor0000}{RGB}{24,24,24}
\definecolor{plantucolor0001}{RGB}{241,241,241}
\definecolor{plantucolor0002}{RGB}{0,0,0}

\lstdefinelanguage{json}{
    basicstyle=\normalfont\ttfamily,
    numbers=left,
    numberstyle=\scriptsize,
    stepnumber=1,
    numbersep=8pt,
    showstringspaces=false,
    breaklines=true,
    frame=lines,
    backgroundcolor=\color{background},
    literate=
     *{0}{{{\color{numb}0}}}{1}
      {1}{{{\color{numb}1}}}{1}
      {2}{{{\color{numb}2}}}{1}
      {3}{{{\color{numb}3}}}{1}
      {4}{{{\color{numb}4}}}{1}
      {5}{{{\color{numb}5}}}{1}
      {6}{{{\color{numb}6}}}{1}
      {7}{{{\color{numb}7}}}{1}
      {8}{{{\color{numb}8}}}{1}
      {9}{{{\color{numb}9}}}{1}
      {:}{{{\color{punct}{:}}}}{1}
      {,}{{{\color{punct}{,}}}}{1}
      {\{}{{{\color{delim}{\{}}}}{1}
      {\}}{{{\color{delim}{\}}}}}{1}
      {[}{{{\color{delim}{[}}}}{1}
      {]}{{{\color{delim}{]}}}}{1},
}

\definecolor{eclipseStrings}{RGB}{42,0.0,255}
\definecolor{eclipseKeywords}{RGB}{127,0,85}
\colorlet{numb}{magenta!60!black}

\newcounter{theo}[section]
\renewcommand{\thetheo}{\arabic{section}.\arabic{theo}}

\definecolor{airforceblue}{rgb}{0.36, 0.54, 0.66}
\definecolor{amber}{rgb}{1.0, 0.75, 0.0}
\definecolor{antiquebrass}{rgb}{0.8, 0.58, 0.46}
\definecolor{bleudefrance}{rgb}{0.19, 0.55, 0.91}
	\definecolor{bittersweet}{rgb}{1.0, 0.44, 0.37}
	\definecolor{bondiblue}{rgb}{0.0, 0.58, 0.71}

\colorlet{mylinkcolor}{bondiblue}
\colorlet{mycitecolor}{bittersweet}
\colorlet{myurlcolor}{bondiblue}

\makeatletter
\newcommand\footnoteref[1]{\protected@xdef\@thefnmark{\ref{#1}}\@footnotemark}
\makeatother

\colorlet{punct}{red!60!black}
\definecolor{background}{HTML}{EEEEEE}
\definecolor{delim}{RGB}{20,105,176}
\colorlet{numb}{magenta!60!black}

\definecolor{gray50}{gray}{.5}
\definecolor{gray40}{gray}{.6}
\definecolor{gray30}{gray}{.7}
\definecolor{gray20}{gray}{.8}
\definecolor{gray10}{gray}{.9}
\definecolor{gray05}{gray}{.95}

\newlength\Linewidth
\def\findlength{\setlength\Linewidth\linewidth
    \addtolength\Linewidth{-4\fboxrule}
    \addtolength\Linewidth{-3\fboxsep}
}

\def\BibTeX{{\rm B\kern-.05em{\sc i\kern-.025em b}\kern-.08em
    T\kern-.1667em\lower.7ex\hbox{E}\kern-.125emX}}
\begin{document}

% \title{Conference Paper Title*\\
% {\footnotesize \textsuperscript{*}Note: Sub-titles are not captured in Xplore and
% should not be used}
% \thanks{Identify applicable funding agency here. If none, delete this.}
% }
% \title{Microservice Semantic Dependency Matrix}
\title{Semantic Dependency in Microservice Architecture}

% The Microservice Dependency Blueprint: A Semantic Perspective

% Microservice Semantic Dependency Blueprint

% Semantic Dependency in Microservice Architecture

% Semantic Dependency Detection in Microservice Architecture

\author{\IEEEauthorblockN{1\textsuperscript{st} Amr S. Abdelfattah}
\IEEEauthorblockA{\textit{SIE, The University of Arizona} \\
% \textit{name of organization (of Aff.)}\\
Tucson, AZ, USA \\
amrelsayed@arizona.edu}
\and
\IEEEauthorblockN{2\textsuperscript{nd} Kari E Cordes}
\IEEEauthorblockA{\textit{SIE, The University of Arizona} \\
% \textit{name of organization (of Aff.)}\\
Tucson, AZ, USA \\
karicordes@arizona.edu}
\and
\IEEEauthorblockN{3\textsuperscript{rd} Austin Medina}
\IEEEauthorblockA{\textit{SIE, The University of Arizona} \\
% \textit{name of organization (of Aff.)}\\
Tucson, AZ, USA \\
austinmedina@arizona.edu}
% \and
% \IEEEauthorblockN{}
% \IEEEauthorblockA{} \\
% \textit{name of organization (of Aff.)}\\
% Tucson, AZ, USA \\
% tcerny@arizona.edu}
\and
\IEEEauthorblockN{\hspace{7cm}4\textsuperscript{th} Tomas Cerny}
% \hspace{8.2cm}
\IEEEauthorblockA{\hspace{7.5cm}\textit{SIE, The University of Arizona} \\
% \textit{name of organization (of Aff.)}\\
\hspace{7.5cm}Tucson, AZ, USA \\
\hspace{7.5cm}tcerny@arizona.edu}
% \and
% \IEEEauthorblockN{5\textsuperscript{th} Given Name Surname}
% \IEEEauthorblockA{\textit{dept. name of organization (of Aff.)} \\
% \textit{name of organization (of Aff.)}\\
% City, Country \\
% email address or ORCID}
% \and
% \IEEEauthorblockN{6\textsuperscript{th} Given Name Surname}
% \IEEEauthorblockA{\textit{dept. name of organization (of Aff.)} \\
% \textit{name of organization (of Aff.)}\\
% City, Country \\
% email address or ORCID}
}
% \author{
% \IEEEauthorblockN{4\textsuperscript{th} Tomas Cerny}
% \IEEEauthorblockA{\textit{SIE, The University of Arizona} \\
% % \textit{name of organization (of Aff.)}\\
% Tucson, AZ, USA \\
% tcerny@arizona.edu}}
% \author{\IEEEauthorblockN{1\textsuperscript{st} Given Name Surname}
% \IEEEauthorblockA{\textit{dept. name of organization (of Aff.)} \\
% \textit{name of organization (of Aff.)}\\
% City, Country \\
% email address or ORCID}
% \and
% \IEEEauthorblockN{2\textsuperscript{nd} Given Name Surname}
% \IEEEauthorblockA{\textit{dept. name of organization (of Aff.)} \\
% \textit{name of organization (of Aff.)}\\
% City, Country \\
% email address or ORCID}
% \and
% \IEEEauthorblockN{3\textsuperscript{rd} Given Name Surname}
% \IEEEauthorblockA{\textit{dept. name of organization (of Aff.)} \\
% \textit{name of organization (of Aff.)}\\
% City, Country \\
% email address or ORCID}
% \and
% \IEEEauthorblockN{4\textsuperscript{th} Given Name Surname}
% \IEEEauthorblockA{\textit{dept. name of organization (of Aff.)} \\
% \textit{name of organization (of Aff.)}\\
% City, Country \\
% email address or ORCID}
% \and
% \IEEEauthorblockN{5\textsuperscript{th} Given Name Surname}
% \IEEEauthorblockA{\textit{dept. name of organization (of Aff.)} \\
% \textit{name of organization (of Aff.)}\\
% City, Country \\
% email address or ORCID}
% \and
% \IEEEauthorblockN{6\textsuperscript{th} Given Name Surname}
% \IEEEauthorblockA{\textit{dept. name of organization (of Aff.)} \\
% \textit{name of organization (of Aff.)}\\
% City, Country \\
% email address or ORCID}
% }

\maketitle

\begin{abstract}
% \label{abstract}

% Microservices have been recognized for over a decade. They reshaped system design enabling decentralization and independence of development teams working on particular microservices. While loosely coupled microservices are desired, it is inevitable for dependencies to arise. However, these dependencies often go unnoticed by development teams. These dependencies could be identified syntactically, however, it's even more challenging when it comes to semantics. When multiple microservices share similar logic between each other. As the system evolves, making changes to one microservice may trigger a ripple effect, risking the consistency of the system, necessitating adjustments in dependent microservices, and increasing maintenance and operational efforts. Tracking different types of dependencies across microservices becomes crucial in anticipating the consequences of development team changes. 
% This paper introduces the Semantic Dependency Matrix (SDM) as an instrument to address this challenge from the point of semantics. We present an automated approach for tracking these dependencies and demonstrate their extraction through a case study.

Microservices have been a key architectural approach for over a decade, transforming system design by promoting decentralization and allowing development teams to work independently on specific microservices. While loosely coupled microservices are ideal, dependencies between them are inevitable. Often, these dependencies go unnoticed by development teams. Although syntactic dependencies can be identified, tracking semantic dependencies — when multiple microservices share similar logic — poses a greater challenge. As systems evolve, changes made to one microservice can trigger ripple effects, jeopardizing system consistency and requiring updates to dependent services, which increases maintenance and operational complexity. Effectively tracking different types of dependencies across microservices is essential for anticipating the impact of such changes.
This paper introduces the Semantic Dependency Matrix as an instrument to address these challenges from a semantic perspective. We propose an automated approach to extract and represent these dependencies and demonstrate its effectiveness through a case study. This paper takes a step further by demonstrating the significance of semantic dependencies, even in cases where there are no direct dependencies between microservices. It shows that these hidden dependencies can exist independently of endpoint or data dependencies, revealing critical connections that might otherwise be overlooked.

% Semantic Dependency Matrix (SDM)

\end{abstract}

\begin{IEEEkeywords}
microservice, dependency, coupling, semantic clones, dependency matrix
\end{IEEEkeywords}

\section{Introduction}
\label{Introduction}

%\todo[inline]{Amr: An introduction about microservice and its components.}

% Without understanding distributed services' dependencies on one another, one simple change in service could have a ripple-down effect due to a chain of coupled function calls. Dependencies can come in the form of shared data objects between services, similar endpoint dependencies, or more complicated logical and implicit dependencies. Implicit dependencies are very difficult to quickly visually spot in code, unlike shared data objects or endpoints, making them cumbersome to identify and keep track of in a large-scale system. A change in one logical piece of a service could produce incompatibility with other services without the developer ever knowing. Being aware of the implicit dependencies within microservices as they evolve can allow practitioners to build better systems and solve critical issues faster.

Microservice architecture has become the mainstream approach for cloud-native systems due to its modular and elastic nature and is increasingly vital for managing the complexity of modern systems~\cite{10.1145/3183628.3183631}. By breaking down the functionalities of complex systems into smaller, independent, and loosely coupled units, microservices enable scalability~\cite{torvekar2019}. However, as these systems grow in complexity and scale, managing the dependencies between services becomes increasingly challenging. While microservices are designed to be loosely coupled, they often exhibit both explicit dependencies—such as direct endpoint calls—and implicit dependencies, including shared logic or semantic relationships, which can lead to inconsistencies across the system~\cite{ghofrani2018}.

\noindent\textbf{Problem Statement.}
Dependencies in microservices arise when changes in one part of the system necessitate corresponding changes in another, whether these dependencies are explicit or implicit. The implicit dependencies, particularly those arising from shared logic, or semantic clones, are especially difficult to detect and manage. Semantic clones occur when two code fragments perform the same function, while the fragments could be syntactically different \cite{kumar2021}. Such clones can create hidden dependencies across microservices, leading to maintenance challenges and potential system inconsistencies. If semantic dependencies go undetected, even small changes in one service can have ripple effects, leading to functionality inconsistency and increased maintenance costs~\cite{kumar2021}.

Microservice architectures, with their distributed nature and component-based structure, complicate the task of identifying these semantic dependencies. When components within or across microservices rely on similar logic or shared functionality, a change in one component can affect others in unpredictable ways. For instance, two microservices may implement a shared flow to retrieve specific data, and if one microservice's implementation is modified without updating the other, it can lead to inconsistent system behavior. This scenario is common when microservices rely on a shared flow but access an endpoint or data source from another microservice or external service to fulfill the process. In such cases, endpoint and data dependency analysis would fail to capture the relationship since there is no direct interaction between the two microservices. However, semantic dependencies effectively reveal this hidden relationship. Without a comprehensive view of these interactions, identifying such issues becomes challenging, especially when system documentation fails to keep up with the evolving complexity of the implementation.
% and complicated details
% For example, two microservices may implement a shared flow to retrieve specific data, and a modification to this flow in one service without the other may lead to inconsistent behavior. It is common in this case of two microservices implemts a common flow, however they use an endpoint or data source from another microservice or external service. In such cases, the dependency is not present in terms of endpoint or data relationships since there is no direct interaction between the two services. However, the semantic dependency effectively captures this indirect relationship, revealing dependencies that would otherwise go unnoticed. Thus, Without a comprehensive understanding of these interactions, detecting such issues is challenging, especially as documentation may not keep pace with evolving system complexity and with such details in the system implementation.

\noindent\textbf{Objective.}
The objective of this paper is to introduce an automated method for detecting semantic dependencies between microservices based on their architectural components. Our methodology focuses on identifying component-level semantic relationships between microservices to identify implicit dependencies that could otherwise go unnoticed. It detects semantic clones between components, meaning that a change in one component necessitates consideration of the other cloned component. This establishes semantic dependencies between the microservices containing those cloned components. By automating the detection of these dependencies from the system source code, we aim to provide practitioners with a holistic view of their system's semantic structure, enabling them to better manage the risks associated with code inconsistencies and maintenance. This paper provides a detailed methodology to achieve semantic dependency detection and visualization, supported by a case study for evaluation. Additionally, the case study explores the potential relationship between semantic dependencies and endpoint and data dependencies, providing a more comprehensive view of the system and offering multiple perspectives for further consideration.

% the case study explores the potential relationship between semantic dependencies on one side and endpoint and data dependencies on the other side, offering insights into a comprehensive picture of multiple perspectives of system and offering different perspectives for further consideration.

% This paper details a methodology for achieving that and execute a case study for evaluation. Moreover, we consider into our case study execution to discuss the potential relation if any between semantic dependencies in one side and endpoint and data dependencies. That should open different aspects of considerations.

\noindent\textbf{Contributions.}
%This paper makes the following key contributions:
The following contributions are made:
\vspace{-0.2em}
\begin{itemize}[leftmargin=*]
\item An automated methodology for detecting component-based semantic dependencies between microservices.

\item A matrix-based visualization that offers a holistic view of semantic dependencies across microservices.

\item A case study and prototype implementation to evaluate the methodology using a well-established benchmark for validation. It showcases the extraction and visualization of semantic dependencies across microservices.

% It demonstrates the extraction and visualziation of semnatic depdenciec accorss microservices in a system.

% \tomas[]{I suggest to say case study demonstrating X..Y..}

\item An exploration point of the integration and relationships among endpoint, data, and semantic dependencies, and providing a holistic perspective on system dependencies.

% A discussion point of the integration and relatinship between different dpendencies of endpoint, data, and semantic dependencies in the holistic perspective of the system.

% \tomas[inline]{I suggest to find discussion point and add it - what is the finding}

\end{itemize}

\noindent\textbf{Paper Organization.} The paper is structured as follows: Section~\ref{sec:background} reviews related work. Section~\ref{sec:methodology} presents the proposed methodology and its phases for constructing the Semantic Dependency Matrix. In Section~\ref{sec:case-study}, we demonstrate the methodology's implementation on a benchmark system. Section~\ref{sec:discussion} elaborates on the methodology's insights and addresses threats to validity. Finally, Section~\ref{sec:conclusion} concludes the paper and summarizes its key contributions.

% The rest of the paper flows as follows: Section~\ref{sec:background} discusses the background information relating to this paper and the related work. Section~\ref{sec:methodology} details the proposed methodlogy and its phases to create the Semantic Dependency Matrix; next section~\ref{sec:case-study} covers a demonstration of our methodology implemenation on a benchmark system; then the section~\ref{sec:discussion} explains further details about the methodlogy directions and mention threats to validity. Finally, Section~\ref{sec:conclusion} concludes the paper and its contributions.

\section{Related Work}
\label{sec:background}

% \todo[inline]{Amr: Background about semantic dependency if needed after writing the introduction.}

% \todo[inline]{Amr: Related work about different considering of semantic dependencies in monolithic systems and microservice systems, show how our approach is different}

% One of the most important considerations in software development is finding and managing dependencies in code \cite{EsparrachiariDependencies}. Other papers \cite{kumar2021, higo2008, svacina2022, Savic2014} have discussed methods of finding semantic clones, but this is a difficult task made more so by the distributed architecture of microservices systems. Semantic clones are also much harder to detect than syntactic clones - code clones that occur when two code fragments have the same or very similar syntax\cite{Abdelfattah2023}. Two semantic clones don't necessarily have the same code, and could be classifed in some studeis as a code smell~\cite{Abdelfattah2023}

One of the critical aspects of software development is identifying and managing code dependencies~\cite{EsparrachiariDependencies}. While other studies~\cite{kumar2021, higo2008, svacina2022, Savic2014} have explored methods for detecting semantic clones in software systems, this task is particularly challenging in microservices-based systems due to the distributed nature of the architecture. Unlike syntactic clones, which are code fragments with similar syntax, semantic clones do not necessarily share the same code structure and can sometimes be classified as code smells in certain studies~\cite{Abdelfattah2023}.

Zhong et al. \cite{ZHONG2023111670} state in their work that it is difficult to manage the quality of microservice applications due to their separated and self-contained nature. Still, it is important to know how coupled the different microservices are for development and rework purposes. Apolinario et al.~\cite{Apolinario2021} highlight the operational complexity of microservices, worsened by limited monitoring tools and disorganized development, leading to high coupling. Studies \cite{panichella2021structuralcouplingmicroservices, ZHONG2023111670, Apolinario2021} emphasize that managing dependencies, particularly semantic ones, is crucial for maintaining loose coupling in microservice applications, motivating this study.

% Apolinario et al. \cite{Apolinario2021} agree that microservices come with increased operational complexity and that they are made even more complex due to it being an emerging architectural style, there are fewer monitoring and metrics that can help make maintenance easier. Moreover, the lack of organization in the development process can lead to high coupling, causing difficulties later on in the development and maintenance of the application. These studies \cite{panichella2021structuralcouplingmicroservices, ZHONG2023111670, Apolinario2021} reveal that managing dependencies is one of the most important considerations in software development. Managing semantic dependencies is vital for maintaining low, loose coupling in microservice applications, motivating this study.
%Finding and managing the semantic dependencies is crucial to ensuring the low and loose coupling of a microservice application, which is why we choose to examine this topic in this work.

% \todo[inline]{and this}
Most code clone detection methods focus on syntactic clones since semantic clones are difficult to handle \cite{Huang2022}. Semantic clones require a more in-depth analysis to identify and detect the meaning behind the code and its structure \cite{Mastroeni2008}. Multiple methods are introduced for clone detection as described in~\cite{Abdelfattah2023, kumar2021, higo2008}. However, most of them are built for monolithic systems. Savic et al.~\cite{Savic2014} proposed a language-independent approach for monolithic systems, using enriched concrete syntax trees (eCSTs) to extract semantic dependencies. However, it focuses solely on monolithic architectures, ignoring microservice systems. In contrast, Zhao et al.~\cite{zhao2022} focused on microservices, using Nicad to detect syntactic clones and a custom tool to analyze cross-service clones, emphasizing the reasons behind these clones.

Moreover, there are significant shortcomings in tools to detect semantic dependencies in microservices. In~\cite{Abdelfattah2023}, Component Call Graphs (CCGs) were introduced as a possible alternative to Control Flow Graphs (CFGs). A CCG is created from the CFG of a code base, where the CFG is condensed into method units so the flow of method calls is shown. The CCG describes the control flow of the system across components, including calls within and between microservices. Using CCGs is much less expensive due to the reduction of low-level program detail. They utilized the CCGs to identify clones across microservices-based systems. However, they did not consider semantic clones under the context of dependencies or address them in the context of other types of dependencies. Nevertheless, we utilized their recommended representation of CCGs to build our methodology accordingly. We provide a visualization for the holistic view of the semantic dependencies in microservices-based systems while considering the component structure and interactions.

Additionally, various visualization approaches have been developed to represent microservice architectures and their dependencies \cite{benomar2013, abdelfattah2023MDM, 10.1145/3538641.3561493, cerny2024static, 10.1145/3477314.3506963, 10174165, 9912621, 9912633}. Specifically related to the presented visualization in this study, Abdelfattah and Cerny \cite{abdelfattah2023MDM} focus on endpoint and data dependencies within microservices-based systems. They introduced heat maps to visualize these dependencies in a matrix format, creating two matrices: the Endpoint Dependency Matrix (EDM) and the Data Dependency Matrix (DDM). These matrices identify which microservices rely on endpoints from another microservice, and which share common data entities with each other. In this study, we apply matrix visualization to our detected semantic dependencies, and in the discussion, we integrate our findings with their matrices, offering a comprehensive view that encompasses all three dependency aspects for further investigation.

This paper presents an automated approach for defining, extracting, and visualizing semantic dependencies within the source code of microservices-based systems. It highlights the significance of recognizing these dependencies for effective system maintainability.

\section{Component-based Semantic Dependency (CSD)}
\label{sec:methodology}
% Methodology
This section outlines the proposed methodology for extracting semantic dependencies from microservice-based systems and representing them in a semantic dependency matrix.

\begin{figure*}[!ht]
\vspace{-1.5em}
    \centering
    \includegraphics[width=0.85\textwidth]{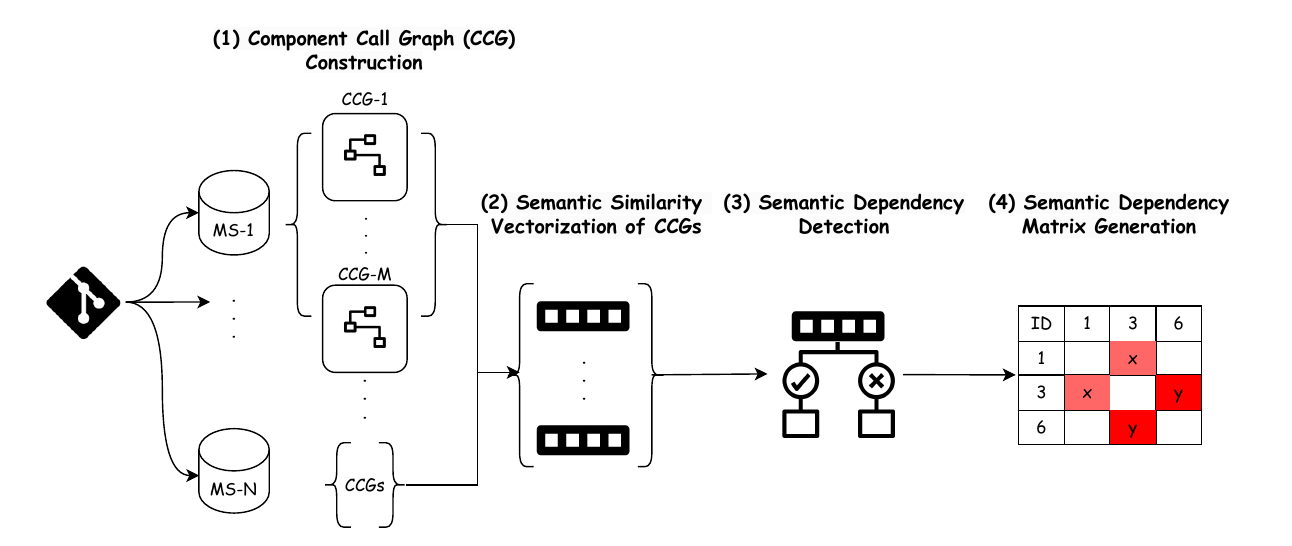}
    \vspace{-0.5em}
    \caption{Component-based Semantic Dependency Detection Process.}
    \label{fig:process}
    \vspace{-2em}
\end{figure*}

\subsection{Methodology Overview}
\label{methodology overview}

% This methodology is Component-based Semantic dependency (CSD). The objective of the methodldogy is to identify the semantic dependecnies between microservices and each other in a system through the focusing on the architectural components' semnatics and their communication flows in the system.

% The CSD addresses the semantic dependencies in microservices-based systems while it focuses on the archtecture components and their communication flow in the system. The CSD builds its technique on the semantic clones and similarities between component call graphs (CCGs). These CCGs represent the main architectural components of the system, including internal calls between different architectural layers, as well as inter-service communications, such as endpoint calls between microservices, as demonstrated in~Figure~\ref{fig:CCG}. The CSD method is useful in its ability to abstract beyond low-level syntactic code and focus more on system endpoints and functional groups that span the architectural layers of a system.

% Our methodology is outlined in~Figure~\ref{fig:process}. It starts by processing the system's source code and constructing CCGs for each microservice. Next, semantic similarities between the CCGs of different microservices are measured. Finally, a heat map matrix is used to visualize the identified semantic dependencies based on the similarities between the CCGs of each pair of microservices in the system.

This methodology, called Component-based Semantic Dependency (CSD), aims to identify the semantic dependencies between microservices within a system by focusing on the semantics of architectural components and their communication flows. Its goal is to offer practitioners a holistic view visualization that facilitates understanding of these semantic relationships as the system evolves, ensuring maintainability and consistency across different microservices in a system.

% It aims to provide the practitioners with a holistic view visualization to make feasible to understand these semantic relaytionship while the system evolves. So, it keeps the system maintanable and consistent from that perspective.

% \tomas[inline]{should you remind the leader what is the goal to get, what problem to solved..? "To provide developers with.. out methodology XYZ" }

% CSD addresses the semantic dependencies in microservice-based systems by emphasizing the architecture components and their interactions. 

The CSD methodology builds on the concept of semantic clones and similarities between CCGs. These CCGs represent the system's architectural components, including internal layer calls, microservice endpoint calls, and interactions beyond the system boundary~\cite{Abdelfattah2023}. The strength of the CSD method lies in its ability to move beyond low-level syntactic code, focusing instead on components interactions that span across architectural layers. Considering the information in CCGs provides a semantic representation of the system's execution flow, capturing interactions between architectural components and making it valuable for understanding the functional and communication relationships within a specific slice of a project.

% These CCGs represent the system's architectural components, including internal calls across architectural layers and endpoint calls across microservices and out the system boundry~\cite{Abdelfattah2023}.

% , as depicted in Figure~\ref{fig:CCG}.\todo{Clarify this sentence doesn't make sense, referenced fig 2 before fig 1}

% , either it is inter-service calls between microservices or external calls

% system endpoints and functional groups

Our methodology, outlined in Figure~\ref{fig:process}, begins by analyzing the system's source code to construct CCGs for each microservice. Then, it measures semantic similarities between the CCGs of different microservices. Finally, a heat map matrix is utilized to visualize the detected semantic dependencies, based on the similarities between the CCGs of each pair of microservices in the system.

% It considers architectural components, their proberties and their interacions to provide a descriptive unit representation that can represent the semantic using both their properties and their communication flow in the system.

% It begins with an analysis of the system's source code, followed by the construction of an intermediate representation in the form of a component call graph. This graph represents the main components of the system and its internal and inter-service communications. Next, we measure the semantic similarities between these component call graphs across different microservices. Finally, we visualize the detected semantic dependencies based on the similarities between the component call graphs of each microservice.

\begin{figure}[h!]
    \centering
    \includegraphics[width=0.7\linewidth]{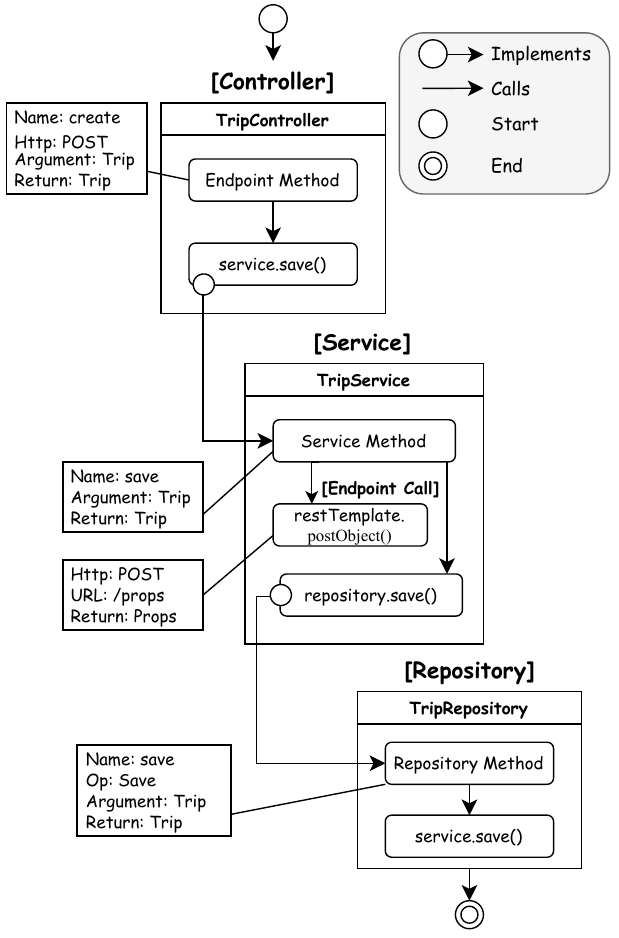}
    \vspace{-0.8em}
    \caption{CCG From Listing 1.}
    \label{fig:CCG}
    % \vspace{-1.2em}
    \vspace{-1em}
\end{figure}

\subsection{Component Call Graph (CCG) Construction (Phase 1)}

Following the standard enterprise layered architecture, the system architecture of each microservice project is divided into components and layers communicating with each other while preserving their single responsibility, such as Controller (handling external communications through endpoints), Service (containing business logic), and Data layer (managing data storage)~\cite{9737496, Abdelfattah2023}. The CCG captures all the necessary information to describe components' properties and interactions in such architecture, as highlighted in~\cite{Abdelfattah2023}. In this methodology, the CCG serves as the granularity level for semantic comparisons. It is as an intermediate representation that shows components' properties, internal component layer calls, and inter-service communications between microservices.

\label{component call}
\begin{table}[!h]
    \centering
    \caption{Component types and their properties.}
    \begin{tabular}{ll}
        \hline
        \textbf{Component Type} & \textbf{Properties} \\[3pt] \hline
        Controller              & Method name, HTTP method, arguments, return type \\[2pt]
        Service                 & Method name, arguments, return type \\[2pt]
        Repository              & Database operation, arguments, return type \\[2pt]
        Endpoint Call               & URL, HTTP method, return type \\[2pt] \hline
    \end{tabular}
    \label{table:component_types}
    \vspace{-2.5em}
\end{table}

To construct CCGs from each microservice's codebase, we applied Software Architecture Reconstruction (SAR)~\cite{rademacher2020modeling, app13031838, 9912633}, a process that extracts system information to create architectural views and analyze the system's structure. For this purpose, we developed a language-specific parser to extract the necessary components and their properties for building CCGs, in line with language standards. Our methodology classifies components into four types: Controller, Service, Repository, and Rest Call, each with specific properties, detailed in Table~\ref{table:component_types}.

Our SAR process involves three main steps. First, we identify the layer-based components, such as \texttt{Controller}, \texttt{Service}, and \texttt{Repository}. For example, In Java Spring, these components are recognized through annotations like \texttt{@Controller}, \texttt{@Service}, and \texttt{@Repository}. Second, we extract the method definitions, including both the method declarations and method bodies, within each component. Third, starting from the top layer (\texttt{Controller}), we identify the defined endpoint methods in each Controller component (each of which serves as the starting point for a CCG) and analyze their bodies to detect method calls to other components such as service method call or endpoint call. We use a signature matching technique to map the extracted method calls to their corresponding method declarations. We adopted a similar technique that arises in profiling systems through log analysis, where the generated log lines match the corresponding log statements in the source code \cite{zhao2014lprof}. %We adopt the same technique as Zhao et al. \cite{zhao2014lprof}. 
Our signature matching method extracts code call statements and builds templates using regular expressions to match parameter types with their corresponding in method declarations.
% Our signature matching method identifies all code call statements to extract templates that can be matched using regular expressions, allowing us to align parameter types with their corresponding in method declarations.

% We follow the same technique of Zhao et al.~\cite{zhao2014lprof}. Our signature matching technique identify all code call statements to extract templates that can be matched using regular expressions to align parameter types with their corresponding in the method declations.

% A similar challenge has been accounted for when profiling systems using log analysis and matching log lines with logging statements in the source code \cite{zhao2014lprof}. Zhao et al.~\cite{zhao2014lprof} have identified all code log statements to extract templates that could be matched using regular expressions to identify and match the parameter types whose values are present in the log output. 

% prioritizing matching order as discussed in~\cite{mention-one}\todo[]{fix}.
% \vspace{-1em}
%\begin{lstlisting}[basicstyle=\footnotesize\ttfamily]
\begin{lstlisting}[language=Java, basicstyle=\footnotesize\ttfamily, caption=Source code example. Note: \texttt{Trip} is a data entity., label=listing:trip_code]
@Controller
public class TripController {
    @Autowired
    private TripService service;

    @PostMapping
    public Trip create(@RequestBody Trip trip) {
        return service.save(trip);
    }
}

@Service
public class TripService {
    @Autowired
    private TripRepository repository;

    public Trip save(Trip trip) {
        Props p = restTemplate.postObject("/props");
        trip.setProps(p);
        return repository.save(trip);
    }
}

@Repository
public interface TripRepository {
    Trip save(Trip trip);
}
\end{lstlisting}
% \vspace{-0.5em}

It is important to note that this phase of constructing CCGs can adopt various approaches to achieve language-agnostic capabilities, as demonstrated in~\cite{9737496}. However, this methodology prioritizes the use of a language-specific parser to ensure a more comprehensive extraction of information, rather than generalizing across multiple languages, which is not the primary focus of this methodology.

For example, as shown in Listing~\ref{listing:trip_code}, an endpoint method \textit{create} in the \textit{TripController} calls the \textit{save} method in the \textit{TripService}, which then invokes both the endpoint \textit{/props} and the \textit{save} method in \textit{TripRepository}. These calls are detected, formulated, and represented in a CCG as illustrated in~Figure~\ref{fig:CCG}. The CCG starts at the controller, the service's endpoint interface, and traces through each method, mapping the entire flow across the architectural layers.

\subsection{Semantic Similarity Vectorization of CCGs (Phase 2)}
\label{semantic detection}

% In this phase, we calculate the semantic similarity between each pair of CCGs from different microservices. As shown in Table \ref{table:component_types}, each component in the CCG has its own proberties. To calculate the semantic similarities between each component and their corresponding on the other CCG. It considers comparing each proberty from that component with the same proberty in the other component. Different methodologies are used to compare each property and assign a similarity score as they each exist in a different context.

This phase involves calculating the semantic similarity between each pair of CCGs from different microservices and generating a semantic similarity vector for each pair. As shown in Table~\ref{table:component_types}, each component in a CCG has its own set of properties. To determine the semantic similarity between components across different CCGs, we compare each property of a component with the corresponding property in the other component. Various methodologies are applied to compare these properties, and a similarity score is calculated.

% , accounting for the differing contexts in which these properties exist. 

% two components we compare their method name, HTTP method, arguments, and return type as given in Table \ref{table:component_types}.

% With the Component Call Graph depicting components and their methods across microservices and some deeper information about those methods to give us a better semantic understanding of what each one does, we can compare the CCGs. Comparing the similarity between each CCG enables us to identify elements of a service that are semantically the same as another element of a service, creating a semantic dependency. The process compares the extracted properties of the CCGs after cross-cutting given in Table \ref{table:component_types}, finding similarity values for each CCG compared to others in the microservice, and then the similarity values are classified as clones or not clones using logistic regression. Moving back to our example from the CCG in Figure \ref{fig:process}, we would compare the controller element to every other controller in the microservice. To compare the two components we compare their method name, HTTP method, arguments, and return type as given in Table \ref{table:component_types}. Different methodologies are used to compare each property and assign a similarity score as they each exist in a different context. \todo[]{why is there an extra blank line here?}

In comparing \textit{method names} we utilize a semantic name comparator based on the WordNet project~\cite{christiane2005}. Using the WordNet project allows us to not only compare syntactic spellings of words to see if they are the same word but also to move beyond their spelling into the semantic meaning of the word. Using semantics allows us to apply our method more generically and regardless of the authors we will be able to find methods whose names describe the same functionality. Semantic word comparison can also be applied to the argument names, and names of custom types. 

\begin{figure*}[h!]
    \centering
    \vspace{-2em}
    \includegraphics[width=0.9\textwidth]{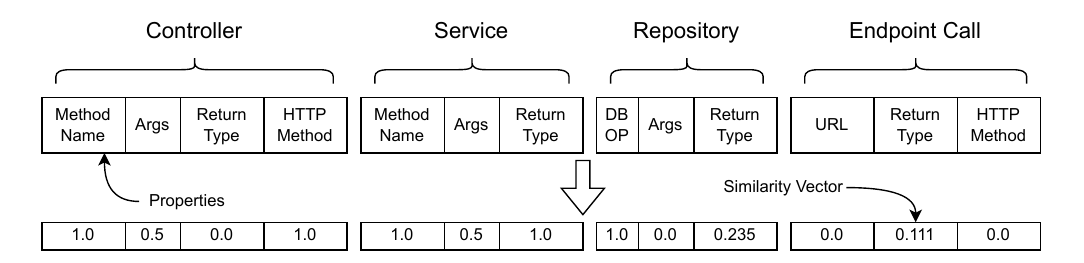}
    \vspace{-0.5em}
    \caption{Components and their Similarity Score.}
    \label{fig:Components}
    \vspace{-1.5em}
\end{figure*}

Next, we consider \textit{data type} comparisons which are seen in the return type and arguments properties. Within data types, there are two options: built-in data types (e.g., String, int) or custom data (e.g., class, struct) types. Built-in data types have always the same name and contain the same type of data so a simple syntactic approach can be utilized to find their similarity. For custom data types, a similar semantic approach used for method names can be implemented. We can start by comparing the variable names of the declared custom types, as well as the names of the custom data types themselves. We can then dive deeper, comparing the names of the variables within the data type. For instance, classes are built primarily using built-in data types so these can be syntactically compared. We can also compare the names of values within a struct to see if they are semantically the same. In instances where the number of variables in a custom data type does not match, the method takes the highest similarity score between combinations of matching between the two custom data types. The final calculated similarity value is the average between all similarity scores for all data types within the specified property.
% a mix of literal and
Building on calculating the similarities, we use syntactic comparisons to find similarities between HTTP methods and URLs. For the \textit{HTTP method}, an exact literal match is necessary as there are only so many HTTP methods, each performing a different function, so if the HTTP method names do not match then the calls do not perform the same semantic function. In terms of the \textit{URL} it can be broken down on %the base and sub-addresses 
the addresses each call travels to and the parameters inserted into the URL. The URLs are compared to each other syntactically, because if a URL goes to a different part of a service than another they will have different URLs, and if the services perform the same semantic function they will have the same parameter types inserted into the call.

The final property to compare is the \textit{database operation}, which is a simple logical matching. If a CCG pair compares a delete with an insert method call, analyzing the query is unnecessary since these functions have entirely different database outcomes. An exception to the literal matching rule occurs as insert and update operations often share semantic effects. In some frameworks, like Spring, a single method may handle both adding a new record (if it doesn't exist) and updating an existing one. Therefore, we treat them as equivalent in our calculations. After calculating property similarities for each CCG pair, we generate a vector representing their component similarity scores, as shown in Figure~\ref{fig:Components}.
% If the database operation in a CCG pair is comparing a delete vs an insert method call, there is no need to analyze the query of what's being inserted or deleted because these two functions have two entirely different outcomes on the database.
% After calculating all of the similarities between properties of each pair of the CCGs, we produce a vector of each CCG pairs, representing the similarity score between their components properties as visualized in Figure~\ref{fig:Components}.
% There is an exception to the literal matching rule as an insert and update operation may use the same method to either add a new record if it is not exist or manipulate the record if it's exist, which have the same effect of adding an effect or maipulating it. so they are considered equal. 

% will have the same end result on the database,

\subsection{Semantic Dependency Detection (Phase 3)}
\label{semantic classification}

% This phase is to detect which of the CCG pair is semnaticlly similar to  cause a depenency between the participating pair of microservices. Using these calculated similarity score vectors from each CCG pairs of the microservices we can analyze them using machine learning classification models to identify whether the calculated similarity vector is consider a similar enough to be as a semantic clone and thus we consider the two microservices containing the participating CCGs having a semantic dependency. 

% Given the potential availability of labeled data for CCG pairs, we propose using supervised machine learning to predict whether two CCGs are similar as a clone based on their similarity vector. We believe these features have descriptive semantic meaning for their CCG and its components, so a simple linear model, such as a logistic regression classifier [36, 37], should be sufficient to capture the relationship. The logistic regression model is a simple yet effective and robust approach that works well with scarce tabular data (as in this usecase). It matches the target of our approach since it calculates a probability based on the different properties of the similarity vector, so it reports if there is a high probability of the CCG pair particopating in the vector values being similar. 

This phase aims to detect which pairs of CCGs are semantically similar enough to indicate cloned semantics between the participating microservices. By utilizing the calculated similarity score vectors from each pair of CCGs, we can analyze them with machine learning classification models to determine whether the similarity vector is sufficiently close to classify the CCGs as semantic clones. If so, we infer that the two microservices containing these CCGs have a semantic dependency. Machine learning is ideal for adapting to the complexity of the patterns and the relationships between CCGs that traditional methods may miss. It offers scalability for large microservices, improving over time with new data.
% and evolves with new data more instances
% Given the potential availability of labeled data for CCG pairs, we propose using supervised machine learning to predict whether two CCGs are similar enough to be considered clones based on their proberties similarity vector. We believe these features of proberties carry descriptive semantic meaning for their respective CCGs and components, making a simple linear model, such as a logistic regression classifier~\cite{bishop2006, james2013}, appropriate for capturing the relationship. 

With the potential availability of labeled data for CCG pairs, we propose applying supervised machine learning to predict whether two CCGs are similar enough to be considered clones based on their calculated property similarity vector. These properties carry semantic meaning for the respective CCGs and components, making a simple linear model, such as a logistic regression classifier~\cite{bishop2006, james2013}, suitable for capturing this relationship. It is further validated through cross-validation~\cite{berrar2019cross} with other models (i.e., decision trees, support vector machines, and random forests). The logistic regression model is effective and robust, particularly when working with limited tabular data, as in this case. It aligns with our approach by calculating the probability based on the various properties of the similarity vector, thus indicating whether there is a high likelihood that the CCG pair shares similar semantics. It is a parametric model that attempts to learn the following function:

% \todo[]{what features - the ones from the vector?}

% For our use case, logistic regression works well because we have existing data of dependencies acquired from manual comparisons of the CCGs, and logistic regression is a form of supervised learning. In logistic regression, we aim to find the set of coefficients that satisfy the following function:
% \todo[]{don't say as stated if it's not the same thing - seems like it's saying the same thing as the sentence before} 
\begin{equation}
    f(x) = \mathbb{P}[y = +1 \mid x]
\end{equation}
where x is the similarity vector, and y is a binary target with y=1 meaning the CCGs pair is semantically similar and y=0 meaning the CCGs are not semantically similar. The goal is to use supervised learning to satisfy equation 1 by finding a set of coefficients in the following equation:
\begin{equation}
    h(x) = \theta(\mathbf{w}^T x + b)
\end{equation}
where w are the coefficients (a vector of scalar weights), $b \in \mathbb{R}$ is the bias term, and $\theta$ is the logistic function defined as:
\begin{equation}
    \theta(z) = \frac{1}{1 + \exp(-z)}, \quad z \in \mathbb{R}
\end{equation}
The model weights and bias terms are adjusted utilizing the gradient descent method \cite{bishop2006}, which aims to minimize the in-sample error rate. Logistic regression models at their core cannot directly tell us whether two CCGs are semantically similar or not, but will rather tell us based on the different properties of the similarity vector if there is a high probability of them being similar \cite{james2013}. The machine learning algorithm assigns weights to similarity scores' properties based on their importance, setting a threshold to classify CCG pairs as semantically similar or not based on the probability.
% The machine learning algorithm will utilize the different similarity scores, assigning different weights to each property depending on how important the algorithm finds each to be in determining similarity. It will then set an internal threshold that if two CCGs are over a certain value, or probability, then we will be confident they can be considered semantically similar, otherwise, we will label them as not similar.
 
\subsection{Semantic Dependency Matrix Generation (Phase 4)}
\label{semantic matrix}

% This phase keep track of each microservice pair and count how many CCG pairs are detected as semanticlly similar in the previous stage. For example, if MS-1 and MS-2 have three differnt pairs of CCG that are classified as semnatically similar, then we consider these two microserrvices have semantic dependencies with the degree of three.

% We visualize the dependencies between each pair of microservices using a heatmap approach, as recommended in \cite{benomar2013}. We convert these semantic dependency pairs into a grid where every microservice with a dependency is listed as a row and a column. Each cell has the value of  total number of CCGs that have sematic similarity on the corresponding two microservices in the row and column. The cell color will be more dark as the number of depenecnies is higher between such pair of microservices.

% Notice that the matrix will be semitric since we consider such semantic dependency is bi-directional.

This phase tracks each pair of microservices and counts how many CCG pairs were identified as semantically similar in the previous stage. For example, if two microservices have three different pairs of CCGs classified as semantically similar, we consider these two microservices to have a semantic dependency with a degree of three.

To represent and visualize the dependencies between each pair of microservices, we employ a heatmap approach, as recommended in \cite{benomar2013,abdelfattah2023MDM}. We convert the semantic dependency pairs into a grid, where each microservice with a dependency is represented as both a row and a column. Each cell in the grid contains the total number of CCGs that exhibit semantic similarity between the two corresponding microservices. The cell color becomes darker as the number of dependencies increases between the pair of microservices.

It is important to note that the semantic dependency is considered bi-directional. This means that if Microservice A has a semantic dependency with Microservice B, then Microservice B also has a semantic dependency with Microservice A, reflecting the mutual nature of their interactions.

\section{Case Study}
\label{sec:case-study}
%\todo[inline]{Amr: Explain the objective of the case study.}
% In this section, we demonstrate the efficacy of our proposed method on an existing benchmark. Using the CSD methodology, we have analyzed an existing microservices-based system for semantic dependencies. We have generated its corresponding Semantic Dependency Matrix, showing which microservices are dependent on each other.

% Our study highlights that the Semantic Dependency Matrix is an important instrument for tracking semantic dependencies between across microservices in a system. Additionally, rerunning the analysis and regenerating the Semantic Dependency Matrix can allow developers to track dependencies in code as they're writing it. Each new commit can be analyzed for semantic dependencies to determine what, if any, dependencies are introduced. or any depenendy is violated causing an inconsistency between microservices and each other.

% \tomas[inline]{what is it we try to prove by the study - that we can extract matrix or to learn the implications? Since we did not get a user study, it is hard to say effectiveness.}

In this section, we evaluate our proposed method using an established benchmark. Using the CSD methodology, we have analyzed an existing microservices-based system to identify semantic dependencies. The semantic interdependencies among the microservices are illustrated in a corresponding Semantic Dependency Matrix.

% This analysis has resulted in the generation of a corresponding Semantic Dependency Matrix, which illustrates the semantic interdependencies among the microservices.

Our study emphasizes that the Semantic Dependency Matrix serves as a valuable instrument for tracking semantic dependencies among microservices within a system. Furthermore, reanalyzing the system and regenerating the Semantic Dependency Matrix enables developers to track dependencies in real-time as they write code. Each new change in the source code can be evaluated for semantic dependencies to identify any new dependencies that are introduced or any violations that may lead to inconsistencies between microservices.

\subsection{Experimental setup}

The experimental setup includes the selection of the benchmark, the data prepared for the execution and as a baseline, and the implementation details of the prototype. A reproducible package containing this produced data and prototype implementation have been published online in the dataset\footnote{\label{foot-dataset}Reproducible Package: \url{https://zenodo.org/records/13927257}}. %, accessed on 10/13/2024

\subsubsection{Benchmark}
\label{benchmark}
%\todo[inline]{Amr: Explain brief about the used benchmark.}

% While we make a pilot validation over a subset of projects from the microservice dataset provided by
% Amoroso et al. \cite{amoroso2023one}. We filtered projects that are based on Java Spring Boot Framework. This dataset mentions TrainTicket\footnote{TrainTicket Version 0.1.0: \url{https://github.com/FudanSELab/train-ticket/releases/tag/v0.1.0} as the largest well-established open-source microservices-based system benchmark, consisting of 41 microservices. Thus, this case study is performed on the TrainTicket, it is a service that allows users to buy train tickets. Communication is facilitated through API calls and a MySQL database is used to store information.

We conducted a pilot validation using a subset of projects from the microservice dataset~\cite{MSR2024}. Among them, TrainTicket\footnote{TrainTicket: \url{github.com/FudanSELab/train-ticket/releases/tag/v0.1.0}}, recognized as the largest and most well-established open-source benchmark for microservices, was selected for this case study. It consists of 41 microservices built using Java in Spring Boot framework and MySQL database.
%, facilitates train ticket purchases through API communication and utilizes a MySQL database for data storage.

% \footnote{TrainTicket Version 0.1.0: \url{https://github.com/FudanSELab/train-ticket/releases/tag/v0.1.0}, accessed on 10/10/2024}~\cite{zhou2018benchmarking}. the The benchmark used is the TrainTicket\footnote{TrainTicket Version 0.1.0: \url{https://github.com/FudanSELab/train-ticket/releases/tag/v0.1.0}, accessed on 10/10/2024}~\cite{zhou2018benchmarking}. %We utilized its latest release V1.0.0\footnote{TrainTicket Benchmark: https://github.com/FudanSELab/train-ticket/tree/v1.0.0, accessed on 10/10/2024}. 
% It is a service that allows users to buy train tickets. Communication is facilitated through API calls and a MySQL database is used to store information. This system is a well-established open-source microservices-based system benchmark, consisting of 41 microservices built using Java in Spring Boot framework.

\subsubsection{Dataset Preparation} 
% TrainTicket release 0.1.0\footnote{TrainTicket Version 0.1.0 https://github.com/FudanSELab/train-ticket/releases/tag/v0.1.0}.
% The dataset was created from the TrainTicket benchmark. We extracted the CCGs using our methodology, then we manually labeled them to mark each pair of CCGs with either clones or non-clone based on the inspecting the code base. This data was manually validated to ensure the accuracy of our methodology. 

% The dataset was generated from the TrainTicket benchmark. We applied our methodology to extract the CCGs. Then, we manually labeled each CCG pair as either a clone or non-clone based on a thorough inspection of the codebase. To ensure the reliability of our approach, the dataset was manually validated for accuracy. The analysis of the code and the validation of the clones were done by separate authors to reduce bias in the study. The dataset comprises 27,221 unique entry points, representing distinct pairs of CCGs. The system includes 238 distinct CCGs across 41 microservices. Of these, 64 CCGs were involved in clone relationships, while 174 were not. Through manual analysis, 32 clone pairs were identified from the cloned CCGs.

The dataset used in this study was derived from the TrainTicket benchmark. Our methodology was applied to extract the CCGs. To identify clone relationships between each CCG pair, a semi-automated approach was employed: a machine learning-based model was used to pre-classify and label CCG pairs as potential clones or non-clones based on structural and semantic features. These classifications were subsequently validated through selective manual inspection to ensure accuracy and reliability. The date labels underwent an additional level of manual validation to ensure thorough reasoning and accuracy in the results. To mitigate bias, the manual validation of clone pairs and the initial code analysis were performed by separate authors. The dataset consists of 27,221 unique entry points, representing distinct pairs of CCGs across 238 components distributed over 41 microservices. Of these, 64 CCGs exhibited clone relationships, while 174 did not. The semi-automated method identified 32 clone pairs within the cloned CCGs, enhancing scalability while maintaining rigorous validation standards.
% Comprehensive details will be provided in a forthcoming publication due to space limitations.

% This dataset consists of 27,221 entry points, which are distinct pairs of CCGs. The system consists of distinct 250 CCGs out of the 41 microservices. Of those, there were 71 CCGs that were participated in clones and 179 that were not. These cloned CCGs were manually analyzed and 36 clones (1 clone for each pair of CCGs) were identified.

% From release 0.1.0, there are 41 microservices. The code for these services were broken down into 250 distinct CCGs. Of those, there were 71 CCGs that were participated in clones and 179 that were not. These cloned CCGs were manually analyzed and 36 clones (1 clone for each pair of CCGs) were identified.

% \todo[inline]{Amr - read this }%Amr: It needs to mention how did we extract it and the dataset description. It's very much we say we extacted from V0.0.1 of TrainTicket and we manually validated it. The tasks were decomposed so the authors extracted the data and others validated it. You explain how many CCG and how many clones and non-clones out of them.}
% For this case study, TrainTicket\footnotemark[1] was especially used due to integration with related work for further analysis. TrainTicket\footnotemark[1] is a service for the user to buy train tickets. Communication is facilitated through API calls and a MySQL database is used to store information.
\subsubsection{Prototype Implementation}
\label{prototype}

We developed a prototype as a proof of concept for our CSD methodology. The prototype comprises four key components that correspond to the phases of the methodology:
\textit{CCG Construction (Phase 1):}
This component performs a source code analysis of an enterprise Java application. We employed Java-specific tools, including the Java Reflection library and Javassist~\cite{javassist2020}, to statically analyze our benchmark and construct the CCGs among its source code. The benchmark utilizes the Java Spring framework, which employs annotations such as \textit{@Controller}, \textit{@RestController} for controllers, \textit{@Repository} for repositories, and \textit{@Service} for services. It also employs \textit{RestTemplate} for making remote calls including the inter-service calls between microservices. We extracted these components from the source code. \\
\textit{Semantic Similarity Vectorization of CCGs (Phase 2):}
In this phase, the prototype calculates the semantic similarity between properties of the components in the CCGs. It assesses the semantic meaning of each property as detailed in the methodology description. The prototype uses the WS4J project and WordNet~\cite{christiane2005} to measure the semantic meaning difference between words to allow us to compare properties, such as method names. Following the comparison, we generate a similarity vector for each pair of constructed CCGs from different microservices. \\
\textit{Semantic Dependency Detection (Phase 3):}
The prototype implements a logistic regression model using the scikit-learn library~\cite{sci-kitLearn} in Python. We train this model on the generated training dataset and then it test and execute the model on the benchmark data. The model identifies whether a pair of CCGs is semantically similar, thereby determining the semantic dependencies between the corresponding microservices.\\
\textit{Semantic Dependency Matrix Generation (Phase 4):} 
Using the classified similarities from the machine learning model, we compile a list of microservices that exhibit semantic dependencies and quantify the number of CCGs they share. We then create a heat map to visualize the semantic dependency matrix for our benchmark, with darker colors in the cells indicating a higher number of shared CCGs between the services.

\subsection{Results}
\label{results}
% \todo[inline]{Amr: Show the process steps results as clear as possible.}

% \todo[inline]{Amr: Show the semantic matrix representation and write could of paragraphs to interpret its information.}

This section presents the results obtained from each phase of the methodology when executing the implemented prototype on the selected benchmark.

\subsubsection{CCG Construction (Phase 1)}
% \todo[inline]{Amr - read this}
% \todo[inline]{Amr: How many CCG per microservice. It can be added to Table II as another column.}
% From version 0.1.0 of the TrainTicket benchmark, 

% The first phase of our methodology was used to extract and construct CCGs from the source code. This resulted in 238 CCGs, matched the manually generated in the dataset. These CCGs are divided up among the microservices. The most maximum number of CCGs in a microservice is 21 CCGs that was for \texttt{ts-admin-basic-info-service} microservice. The services with the fewest CCGs were \texttt{ts-preserve-service, ts-preserve-other-service, and ts-verification-code-service} which each resulted in 2 CCGs produced. The most frequent number of CCGs (Mode) is 6, which is exist in 7 different microservices.

The first phase of our methodology extracted and constructed 238 CCGs from the source code, matching the manually generated CCGs in the dataset. These CCGs were distributed across the microservices, with the \texttt{ts-admin-basic-info-service} microservice containing the highest number at 21 CCGs. On the lower end, \texttt{ts-preserve-service}, \texttt{ts-preserve-other-service}, and \texttt{ts-verification-code-service} each had 2 CCGs. The most common number of CCGs per microservice was 6, observed in 7 different microservices.

\subsubsection{Semantic Similarity Vectorization of CCG (Phase 2)}
% \todo[inline]{Amr - read this}

This phase utilized the constructed CCGs from all microservices in the system, generating 27,221 distinct pairs of CCGs across different microservices. The proposed vectorization methodology was then applied to the properties of each CCG pair. This process produced a vector of 13 values for each pair, corresponding to the properties listed in~Table~\ref{table:component_types}. These values represent the similarity degree for each of those respective properties within each CCG pair.

% This phase used the constructed CCGs from all microservices in the system, and it generated the combinations of CCGs pairs among the distinct micrsoervuces. It generated 27,221 distinct pairs of CCGs. Then, it applied the proposed vectorization methodlogy on the differnt proberties in each pair of CCGs. This phase generated a vector of size 13 values per each pair (each per each proberties showed in Table~\ref{table:component_types}).

% For determining which clones are semantic dependencies, we measure the similarity of the CCGs constucted in the last step. To compare a pair of CCGs, the average similarity scores for each component are set from analyzing the code with a language parser. Then the similarity scores are compared using machine learning assigned weights in a regression model to determine if a pair of CCGs are semantic dependencies or not. For the training dataset, which was the CCGs constructed from version 0.1.0 of TrainTicket\footnotemark[1]. 

% There were 27,218 pairs of CCGs total that we compared, and from these, 32 semantic dependencies were found.

% \todo[inline]{Amr: how many pair comparison we executed.}

\subsubsection{Semantic Dependency Detection (Phase 3)}
% \todo[inline]{Amr - read this}
%\todo[inline]{Amr: accuracy metric, confusion matrix, and threshold calculation. The clarification and justification of the numbers are required.}

We applied the logistic regression model on the generated similarity vectors, as detailed in the methodology. Given the imbalance in the dataset, a weighted version of the logistic regression model was utilized to emphasize accurate classification of the undersampled (clone) class while reducing the weight for the oversampled class (non-clone). The prediction threshold was set at 0.99, which optimized the $F_1$ score for the dataset \cite{lemnaru2012imbalanced}. In machine learning, this threshold value is often referred to as a hyperparameter and is typically determined using cross-validation techniques~\cite{abu2012learning}. However, due to the limited data, particularly the scarcity of clone samples in this dataset, cross-validation was not feasible in this case study.

\begin{figure}[h!]
    \centering
    \vspace{-1em}
    \includegraphics[width=0.8\linewidth]{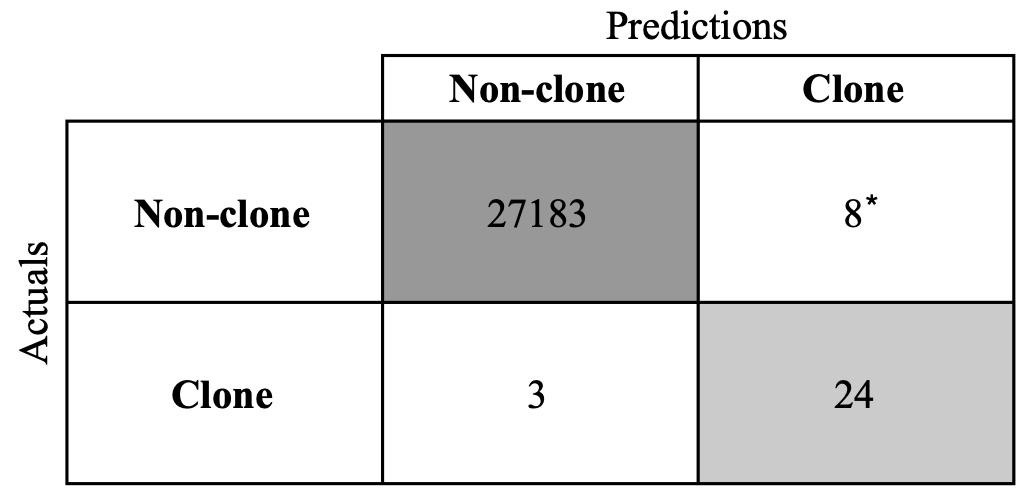}\\
    \raggedright\scriptsize{*: Upon re-examination, 6 of those 8 pairs were found to be true clones that had been mislabeled during manual analysis.}
    \caption{Confusion Matrix for the clones classification in TrainTicket. \small{(Top left is true negatives, top right is false positives, bottom left is false negatives, and bottom right is true positives).}}
    \vspace{-0.7em}
    \label{fig:confusionMatrix}
\end{figure}

The model processed the generated 27,221 similarity vectors, each representing a unique CCG pair. The predicted classifications were compared with the actual labels from the labeled dataset. As shown in Figure~\ref{fig:confusionMatrix}, out of 32 actual clone pairs, the model correctly identified 24. However, 3 pairs were falsely classified as non-clones when they should have been labeled as clones based on manual analysis. Additionally, 8 pairs were incorrectly classified as clones. Upon re-examination, 6 of those 8 pairs were found to be true clones that had been mislabeled during manual analysis, highlighting the model's ability to uncover mistakes in the dataset by learning the component properties.

The three false negatives (semantic clones missed by the model) likely resulted from dissimilar semantics, despite the components performing the same function. In these cases, comparisons involved a REST call versus a repository call performing the same operation. For example, one false negative compared a repository call to delete a contact in \texttt{ts-contacts-service} with a REST call in \texttt{ts-admin-basic-info-service} that performed an identical function.

\noindent\textbf{Metrics Analysis.}
% A summary of the model's performance in the dataset in terms of accuracy, balanced accuracy, precision, recall, and $F_1$ score is given in Table~\ref{tab:log_reg_res}.
The performance of the model on the dataset is summarized in Table~\ref{tab:log_reg_res}, detailing key metrics including accuracy, balanced accuracy, precision, recall, and the $F_1$ score \cite{manning2008introduction, lemnaru2012imbalanced}. Upon analyzing the model's performance, we found it achieved a very high accuracy rate, with most examples correctly classified. However, this metric is somewhat misleading due to the dataset's imbalance. A more reliable measure is the balanced accuracy, which accounts for this imbalance and stands at 94.4\%. The model's precision of 75\% demonstrates that, out of every four examples predicted as clones, one is likely a non-clone. Meanwhile, the high recall indicates that the model is effective at identifying most existing clones. Since the recall is less than 1, it's expected that some actual clones may be misclassified as non-clones. Lastly, the $F_1$ score, which combines precision and recall, confirms the model's overall performance.

\begin{table}[h!]
    \caption{Metrics Measurement Results on the dataset.}
    \vspace{-1.2em}
    \label{tab:log_reg_res}
    % \vskip 0.15in
    \begin{center}
    \begin{tabular}{|c|c|c|c|c|}
        \hline
        \textbf{Accuracy} & \textbf{Balanced Accuracy}  & \textbf{Precision} & \textbf{Recall} & \textbf{$F_1$ Score}\\
        \hline
        0.999 & 0.944 &  0.75 & 0.888 & 0.813\\
        \hline
    \end{tabular}
    \end{center}
    \vspace{-1.5em}
    % \vskip -0.1in
\end{table}

\noindent\textbf{Weights Analysis.}
% The logistic regression model produces weight values that reflect the most important features considered in making predictions. The results of these weights are displayed in Table~\ref{tab:property-weights}. Upon analysis of the weights, it was observed that the controller method name and return type, and the service return type were the most significant factors in the model's predictions.
The logistic regression model generates weight values that indicate the importance of various features in the prediction process. These weight values, which reflect how much each feature contributes to the model's decision-making, are presented in Table~\ref{tab:property-weights}. 

Analyzing these weights shows that the controller method name, the return type of the controller method, and the return type of the service emerged as the most influential factors in the model's predictions. This emphasizes that these particular features play a critical role in determining the semantic similarities between components in the microservices architecture.

\begin{table}[h!]
    \centering
    \caption{Weights for different properties in the similarity vectors. \small{(The bolded ones are the weights with the highest values, contributing most to the similarity score).}}
    % \scriptsize
    \begin{tabular}{|l|l|r|}
        \hline
        \textbf{Component} & \textbf{Property}         & \textbf{Weight} \\ \hline
        \multirow{4}{*}{Controller} & method name      & \textbf{9.06608}         \\ \cline{2-3}
                                   & HTTP method       & 1.38333         \\ \cline{2-3}
                                   & arguments         & 1.86721         \\ \cline{2-3}
                                   & return type       & \textbf{4.39439}         \\ \hline
        \multirow{3}{*}{Service}    & method name      & 1.87319         \\ \cline{2-3}
                                   & arguments         & 0.28254         \\ \cline{2-3}
                                   & return type       & \textbf{4.02174}         \\ \hline
        \multirow{3}{*}{Repository} & database operation & -1.56188      \\ \cline{2-3}
                                   & arguments         & -2.21971        \\ \cline{2-3}
                                   & return type       & 2.91226         \\ \hline
        \multirow{3}{*}{REST Call}  & URL              & -0.58029        \\ \cline{2-3}
                                   & HTTP method       & 1.24877         \\ \cline{2-3}
                                   & return type       & 0.00397         \\ \hline
    \end{tabular}
    \label{tab:property-weights}
\end{table}

% To determine the accuracy of our method, we caluclated it in two ways - the accuracy and the balanced accuracy. The accuracy is calculated as follows:
% \begin{equation}
%     A = \frac{TP + TN}{TP + FN + TN + FP}
% \end{equation}
% and the accuracy for our method on the training set was 0.999. This is due to the number of true negatives so far outweighing the sum of the false negatives and false positives that the ratio is very close to 1. 

% The balanced accuracy is also helpful in this case, since it puts higher weight on correctly classifying an instance of an undersampled category versus correctly identifying an instance of an oversampled one. The balanced accuracy is calculated like so:
% \begin{equation}
%     BA = \frac{1}{2}(\frac{TP}{TP + FN} + \frac{TN}{TN + FP})
% \end{equation}
% The balanced accuracy for this methodology is 0.944. This is less than the accuracy, but considers correctly identifying dependencies more than the accuracy. The first term inside the parentheses wouldn't be extremely close to 1 because this evaluates to 24/(24+3), 0.889. The average of this and 1 essentially would be 0.944. 

\subsubsection{Semantic Dependency Matrix Generation (Phase 4)}

% \todo[inline]{Amr: Change Symetric in methodlogy}

% The Semantic Dependency Matrix is depicted in Figure \ref{fig:Semantic Dependency Matrix} shows for each dependent microservice, the number of CCGs it has in common with each other microservice. The microservice names on the left side and the top side correspond to a specific microservice, and in the intersection of each two microservices is the degree of semantic dependencies in terms of number of CCGs that that microservices pair has in common with each other. Only microservices that shared a dependency with at least one other microservice are included in this matrix. Each microservice name has a number between paranthese indicates the total number of CCGs included in the microservices. The color coding shows darker color when the number of common CCGs are higher comparing to the light color for less number of similar CCGs between the microservices.

% , which is why not every microservice from Table \ref{tab:ms-list} is represented in the matrix (Figure \ref{fig:Semantic Dependency Matrix}).

\begin{figure}[h!]
    \centering
    \vspace{-0.5em}
    \includegraphics[width=1\linewidth]{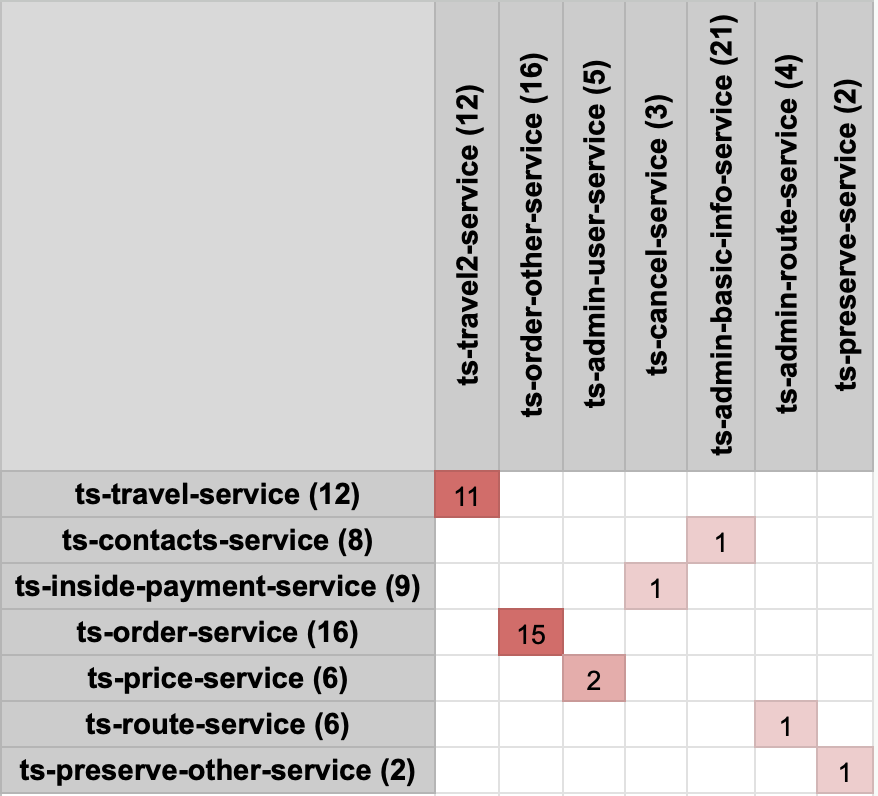}
\caption{Semantic Dependency Matrix of TrainTicket. \small{(The number in parentheses represents the total number of CCGs in each microservice).}}
    \label{fig:Semantic Dependency Matrix}
    \vspace{-0.5em}
\end{figure}

The Semantic Dependency Matrix (Figure~\ref{fig:Semantic Dependency Matrix}) provides a detailed overview of the dependencies between microservices based on their shared CCGs. In this matrix, the microservice names are listed both along the left side and across the top. The intersection of each pair of microservices indicates the degree of semantic dependency, represented by the number of CCGs they have in common. Only microservices that share at least one dependency with another are included in the matrix.

For each microservice, the total number of CCGs it contains is indicated in parentheses next to its name. The color coding in the matrix visually reflects the strength of these dependencies: darker shades represent a higher number of shared CCGs between two microservices, while lighter shades indicate fewer commonalities. This matrix offers a clear, comparative view of the semantic relationships across the system, making it easier to identify how tightly coupled microservices are based on shared functionalities.

% \todo[]{Say in the caption that #CCGs in parentheses - it really looks like the id}

% : Each intersection of two microservices from TrainTicket is marked with the number of how many CCGs they share.

The matrix highlights that the pair of microservices that share the most semantic dependencies (15 CCGs) are \texttt{ts-order-other-service} and \texttt{ts-order-service}. Based on the names, the services appear to be performing very similar if not identical functionality, so it is reasonable for many of the semantic dependencies to be found in these services. The semantic dependency in the code for these microservices is due to multiple components in the code - each service has an API route to get the price of an order that has the same parameters, same return type, and a slight change in the route URL. They also each contain a class to call their respective API route, and these two classes share the same logic except in a few places (database used, class name, and logging information). They each contain a method to access their respective repositories as well. The pair of microservices that share the second most number (11 CCGs) of semantic dependencies are \texttt{ts-travel-service} and \texttt{ts-travel2-service}. These services also share similar API routes, database schema, and other code fragments.

Based on these results, it is crucial to exercise caution when modifying one service in a pair that shares significant semantic dependencies. Changing one service without updating the other can introduce errors that may propagate through the system, affecting even unrelated services. Additionally, this can lead to inconsistencies in business logic across workflows, making it difficult to detect and trace the root cause of such issues. These types of problems can be costly in terms of time, money, and effort to identify and maintain. However, the Semantic Dependency Matrix provides an instrument to depict this perspective in a holistic view. It highlights these hidden dependencies, helping prevent these costly mistakes.

\section{Discussion and Broader Context}
\label{sec:discussion}
% In the proposed method, we aim to demonstrate a method to detect and easily visualize semantic dependencies. The Semantic Dependency Matrix 

The proposed methodology and case study provide a valuable new perspective by extracting semantic dependencies from an architectural components viewpoint, aligning with the microservice-based system design approach. This focus on component-level semantics complements the typical emphasis on endpoint and data dependencies found in much of the literature. Adding semantic dependencies introduces a fresh dimension to understanding system behavior, opening up new research opportunities to explore the relationship between endpoint, data, and semantic dependencies.

\subsection{Merged Dependency Matrix}

\begin{figure*}[!ht]
    \centering
    \includegraphics[width=\linewidth]{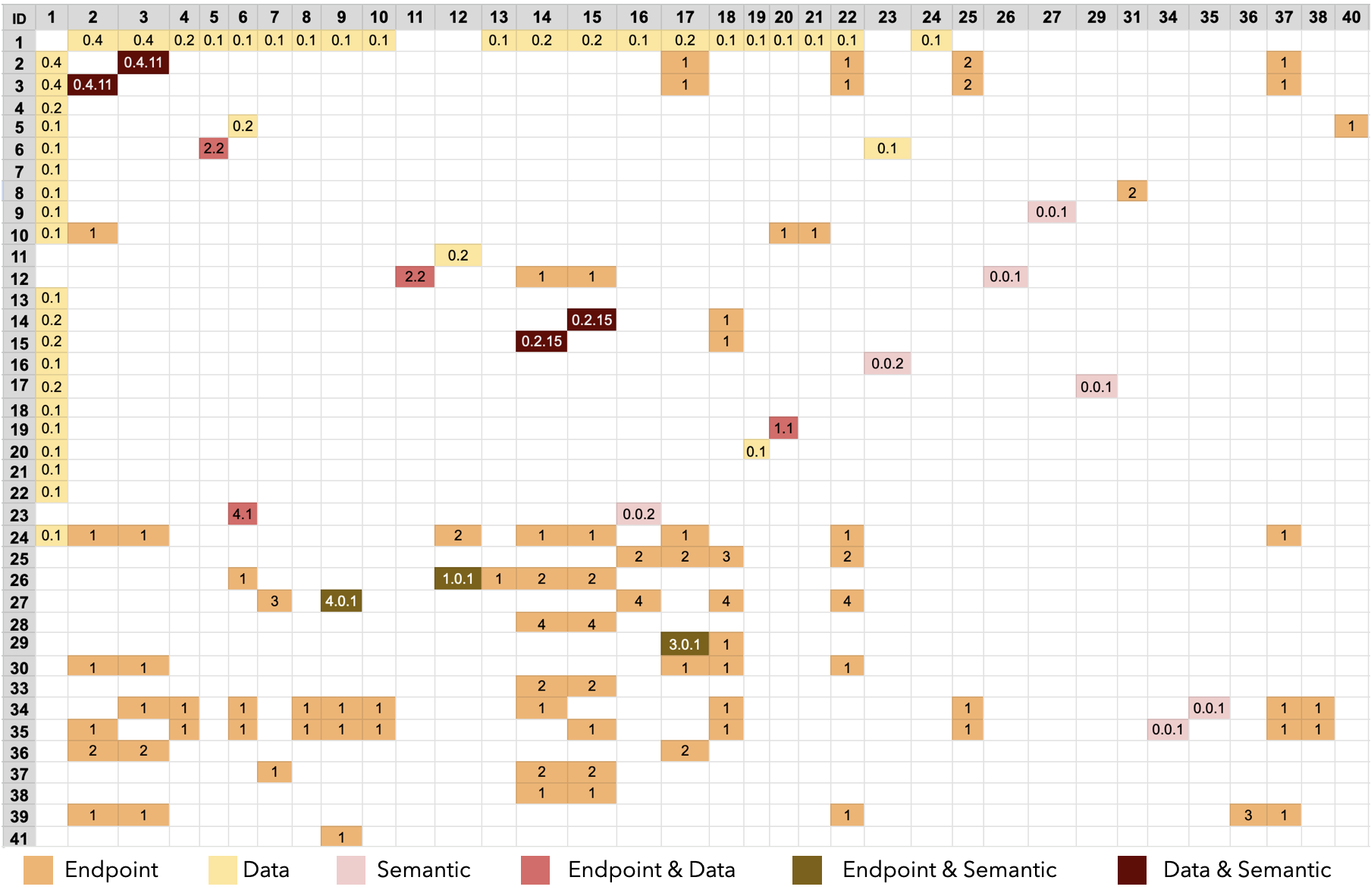}
    \caption{Merged Dependency Matrix - includes Endpoint, Data, and Semantic dependencies.}
    \label{fig:MergedSDM}
    \vspace{-1.5em}
\end{figure*}

To integrate this perspective into the study, we combined our Semantic Dependency Matrix with results from related work on endpoint and data dependencies by Abdelfattah et al.~\cite{abdelfattah2023MDM}. The authors utilized the same TrainTicket benchmark that we used in this study. This allows for a comprehensive view of dependencies across microservices, visualized in the Merged Dependency Matrix (Figure~\ref{fig:MergedSDM}), which includes endpoint, data, and semantic dependencies. The IDs in the top row and left column of the matrix correspond to specific microservices, as listed in Table~\ref{tab:ms-list}. In this matrix, each cell at the intersection of two microservices contains three numbers in the format \texttt{X.Y.Z}. Such that, \texttt{X} represents the number of endpoint calls (unidirectional), \texttt{Y} represents the number of shared data entities (bidirectional), and \texttt{Z} indicates the degree of semantic dependencies (bidirectional).

% \caption{List of TrainTicket microservices from V1.0.0 and their associated IDs}

% \todo[]{don't ever reference this table - add a sentence somewhere. Paragraph right after table?}
\begin{table}[h!]
\centering % Center the table
\caption{List of TrainTicket microservices and their IDs.}
\scriptsize
\begin{tabular}{|l|l|l|l|}
\hline
\textbf{ID} & \textbf{Name} & \textbf{ID} & \textbf{Name} \\
\hline
1 & ts-common & 22 & ts-train-service \\
2 & ts-travel-service & 23 & ts-admin-user-service \\
3 & ts-travel2-service & 24 & ts-rebook-service \\
4 & ts-assurance-service & 25 & ts-basic-service \\
5 & ts-auth-service & 26 & ts-cancel-service \\
6 & ts-user-service & 27 & ts-admin-basic-info-service \\
7 & ts-config-service & 28 & ts-admin-order-service \\
8 & ts-consign-service & 29 & ts-admin-route-service \\
9 & ts-contacts-service & 30 &  ts-admin-travel-service \\
10 & ts-food-service & 31 &  ts-consign-price-service \\
11 & ts-payment-service & 32 & ts-delivery-service \\
12 & ts-inside-payment-service & 33 & ts-execute-service \\
13 & ts-notification-service & 34 & ts-preserve-other-service \\
14 & ts-order-other-service & 35 & ts-preserve-service \\
15 & ts-order-service & 36 & ts-route-plan-service \\
16 & ts-price-service & 37 & ts-seat-service \\
17 & ts-route-service & 38 & ts-security-service \\
18 & ts-station-service & 39 & ts-travel-plan-service \\
19 & ts-food-delivery-service & 40 & ts-verification-code-service \\
20 & ts-station-food-service & 41 & ts-wait-order-service \\
21 & ts-train-food-service & 42 & ts-gateway-service \\
\hline
\end{tabular}
\label{tab:ms-list}
\vspace{-0.7em}
\end{table}

For instance, in the matrix cell at row 3 and column 2, the value \texttt{0.4.11} indicates that between microservices with IDs 2 (\texttt{ts-travel-service}) and 3 (\texttt{ts-travel2-service}), there are 0 endpoint dependencies, 4 data dependencies, and 11 semantic dependencies. Similarly, the cell at row 27 and column 9 shows a value of \texttt{4.0.1}, meaning that \texttt{ts-admin-basic-info-service} (ID 27) made 4 calls to \texttt{ts-contacts-service} (ID 9), with no shared data dependencies and 1 semantic dependency.

To further distinguish between the types of dependencies shared by microservices, we used different color codes: orange for only endpoint dependencies, yellow for only data dependencies, light pink for only semantic dependencies, dark pink for both endpoint and data dependencies, dark yellow-green for endpoint and semantic dependencies, and maroon for data and semantic dependencies. No two services share all three dependency types, so that is not represented with a color.

% There are no two services that share all three types of dependencies, so that is not represented with a color.

\subsection{Different Dependencies Relationship}

The results of the Merged Dependency Matrix demonstrate that many dependencies between services are overlooked by solely considering data or endpoint dependencies. The dependency matrix can help visualize multiple types of dependencies in a code base, and show how important a holistic view of the system can be. From viewing the Merged Dependency Matrix, we can see that there are not that many overlaps between semantic dependencies and other kinds of dependencies. Many of the microservices share only data dependencies or only endpoint dependencies. For example, all the identified dependencies related to \texttt{ts-commmon} (ID 1) are all data dependencies, and none of the services and this service share a semantic dependency. There are also several places where the services share only semantic dependencies, such as \texttt{ts-preserve-other-service} (ID 34) and \texttt{ts-preserve-service} (ID 35). Interestingly, one pair of services discussed with the semantic dependency matrix, \texttt{ts-order-other-service} (ID 14) and \texttt{ts-order-service} (ID 15), are classified as sharing both semantic dependencies and data dependencies. This is due to them sharing a couple of data entities, and containing similar semantics to use that entity.

% Diving deeper into one pair happens in row 23 (\texttt{ts-admin-user-service}) and column 6 (\texttt{ts-user-service}) since they highlight an endpoint and data depndency with value (4.1), however, the don't indicate any semantic similarities.

Diving deeper into the pair in row 12 (\texttt{ts-inside-payment-service}) and column 11 (\texttt{ts-payment-service}), they indicate both endpoint and data dependencies (with value \texttt{2.2}), with values of 2 for each. However, no semantic dependencies are indicated between them. Upon manual inspection, we found that these services share endpoint interactions and two common entities (\texttt{Payment} and \texttt{Money}). Despite this, their architectural semantics are different. Inspecting the \texttt{pay} flow and its related CCGs in these microservices, in the \texttt{ts-payment-service}, it receives calls through an endpoint, validates the data, and then saves it to the database using a repository component. In contrast, the \texttt{pay} flow in \texttt{ts-inside-payment-service} is more complex, interacting with additional components, including making an endpoint call to \texttt{ts-payment-service} to save payment information, as well as calling two other endpoints in \texttt{ts-order-service} and \texttt{ts-order-other-service}. Additionally, it accesses two repositories \texttt{PaymentRepository} and \texttt{MoneyRepository} to retrieve and save data. Our proposed model accurately detects these distinctions, determining that the CCGs of these services are not sufficiently similar to indicate a semantic dependency. This is a reasonable outcome, as while the services share data and endpoint dependencies, their semantic architectures diverge.

Another example can be found in the pair from row 9 (\texttt{ts-contacts-service}) and column 27 (\texttt{ts-admin-basic-info-service}), where the matrix indicates a value of (0.0.1). This shows there are neither endpoint nor data dependencies, but there is a semantic dependency. Upon closer inspection, \texttt{ts-contacts-service} does not initiate any calls to \texttt{ts-admin-basic-info-service}, explaining the absence of endpoint dependencies. Similarly, there are no shared data entities, as \texttt{ts-admin-basic-info-service} typically processes and returns responses without parsing them into corresponding models, and it uses minimal data entities. However, a semantic dependency exists in the \texttt{getAllContacts} flow, and both services have corresponding CCGs for this flow. In \texttt{ts-contacts-service}, data is retrieved via a repository, while \texttt{ts-admin-basic-info-service} accesses it through an endpoint, illustrating two microservices relying on a shared flow through different sources. In such cases, the dependency is not present in terms of endpoint or data relationships since there is no direct interaction between the two services. However, examining semantic dependencies helps to capture hidden, indirect relationships that would otherwise go unnoticed.

% In \texttt{ts-contacts-service}, the data is retrieved through a repository component, whereas in \texttt{ts-admin-basic-info-service}, the data is retrieved via an endpoint call. This case is particularly noteworthy because it highlights a situation where two microservices rely on a common flow, however through an endpoint or data source from another microservice.

% \todo[]{amr look at this sentence}

% However, the semantic dependency effectively captures this indirect relationship, revealing dependencies that would otherwise go unnoticed.

The critical importance of accounting for semantic dependencies becomes evident when examining microservice interactions. Unlike endpoint and data dependencies, semantic dependencies capture deeper, often hidden relationships between services that may not be immediately visible through direct communication or shared data. Overlooking these dependencies during system development or maintenance can result in severe inconsistencies across microservices, leading to unpredictable system behavior, difficult-to-trace bugs, and cascading failures. These overlooked semantic links can disrupt business logic flows, introduce subtle errors, and ultimately increase the complexity and cost of debugging and maintaining the system. Thus, addressing semantic dependencies as a distinct and crucial aspect is vital to ensuring the system's stability and long-term maintainability.

\subsection{Threats to Validity}
% \tomas[inline]{This section is too broad, remove what it describes and describe it only}
\label{threats}

Following Wohlin's taxonomy~\cite{Wohlin2000}, the \textit{Construct Validity} is ensured by utilizing an open-source system developed according to enterprise standards. %The selected benchmark is widely employed in research and recognized as a representative microservice system. The constructed dataset was semi-automated constructed and validated, as highlighted above, the comprehensive validation details can be provided in a forthcoming dataset publication due to space limitations. 
The selected benchmark is widely recognized in research as a representative microservice system. The dataset was constructed and validated using the highlighted semi-automated process and to ensure accuracy. Detailed dataset construction procedures will be published in a forthcoming dataset-focused paper due to space constraints. Regarding \textit{Internal Validity}, in certain cases, the prototype tool might encounter challenges in accurately matching method signatures, particularly when method names are ambiguous. However, we conducted manual validation of the prototype's outcomes to ensure complete information extraction from the source code. For External Validity, our prototype is currently tailored to the Java platform, which may limit direct applicability to other programming languages. However, the methodology itself is designed to be general and adaptable, making it transferable to various languages and frameworks beyond Java and Spring framework. %For \textit{External Validity}, our prototype is designed specifically for the Java platform, which may limit its applicability to other programming languages. It is important to note that the emphasis was on introducing the methodology rather than developing a comprehensive tool. 
Another concern is that the calibration of the weights used for component comparison may not be generalizable. To address this, we utilized a representative microservice application that follows established enterprise standards. Furthermore, machine learning techniques informed the weighting process, although this may be application-specific. To generalize the model, additional training across a wider variety of applications will be necessary. In terms of \textit{Conclusion Validity}, while the benchmark illustrates our case study results, different applications may yield varying outcomes. Additionally, as the benchmark excludes messaging, our findings may not apply to non-component-based systems.

\section{Conclusion}
\label{sec:conclusion}
% \todo[inline]{Amr: Conclude the paper, its takeaway messages, and its future work.}
% Semantic dependencies are a prevalent issue \tomas{issue of what, what scale how important compare to others? all smells are - we identify and study it but the impact remains unknown extend ;)} with software development that can negatively impact the development and maintenance of software. These challenges are exacerbated for microservice systems, due to their distributed nature. Historically, most people have leaned towards data and endpoint dependencies, overlooking semantic dependencies and their implications due to difficulties discovering them. In this work, we introduced a method that can detect semantic dependencies and a tool to clearly visualize them. The Semantic Dependency Matrix helps provide a holistic and scalable view to compare dependencies across microservices and helps developers keep track of these dependencies.

This paper addresses the often-overlooked yet crucial aspect of semantic dependencies in microservice-based systems. Due to the distributed nature of microservices, these dependencies can remain hidden, leading to significant challenges in system maintenance, scalability, and evolution. To tackle this, we proposed a methodology for detecting semantic dependencies directly from the source code, emphasizing a component-based approach that captures the interactions between various application components and their flows.

The presented Semantic Dependency Matrix provides a holistic visualization of these dependencies, offering practitioners a powerful tool to track and manage them effectively. A case study and a developed a prototype are used to evaluate the proposed method using a well-established benchmark. Additionally, by merging semantic dependencies with endpoint and data dependencies, we deliver a comprehensive perspective on microservice interactions, as it is demonstrated in the merged dependency matrix. This combined approach ensures a more thorough understanding of system behavior, ultimately improving system maintainability by helping developers mitigate issues that may arise from hidden dependencies.

Future work will involve conducting an empirical study with practitioners to assess the practical significance of semantic dependencies in real-world systems. Further research will explore the relationships between different dependency types.
% to identify potential correlations or causal links based on system attributes.
% further research will explore the relationships between different types of dependencies (endpoint, data, semantic, and more) to determine if any correlations or causal links can be generalized based on system attributes.

\noindent{\textit{\textbf{ACKNOWLEDGMENT.}}}
This material is based upon work supported by the National Science Foundation (NSF) under Grant No. 2409933 and Grant No. 2229385. %The opinions, findings, and conclusions in this material are those of the authors and do not necessarily reflect the NSF's views.

\bibliographystyle{IEEEtran}
\bibliography{references}

\end{document}